%% file: main.tex
\documentclass[conference]{IEEEtran}
\IEEEoverridecommandlockouts

\usepackage{cite}
\usepackage{amsmath,amssymb,amsfonts}
\usepackage{graphicx}
\usepackage{textcomp}
\usepackage{xcolor}
\usepackage{booktabs}
\usepackage{url}

\title{An Interpretable, Controllable Time-Varying IIR Denoiser for On-Device Assistive Hearing}

\author{
\IEEEauthorblockN{Riccardo Rota}
\IEEEauthorblockA{\textit{Logitech Europe S.A.}, Switzerland \\
\textit{EPFL (\'Ecole Polytechnique F\'ed\'erale de Lausanne)}, Switzerland \\
rrota@logitech.com \\
ORCID: 0009-0003-2952-0421}
\and
\IEEEauthorblockN{Kiril Ratmanski}
\IEEEauthorblockA{\textit{Logitech Europe S.A.}, Switzerland \\
kratmanski@logitech.com \\
ORCID: 0000-0002-1763-3459}
\and
\IEEEauthorblockN{Jozef Coldenhoff}
\IEEEauthorblockA{\textit{Logitech Europe S.A.}, Switzerland \\
jcoldenhoff@logitech.com \\
ORCID: 0009-0005-1669-2579}
\and
\IEEEauthorblockN{Milos Cernak}
\IEEEauthorblockA{\textit{Logitech Europe S.A.}, Switzerland \\
mcernak@logitech.com \\
ORCID: 0000-0002-5569-9491}
}

\begin{document}

\maketitle

\input{0_abstract.tex}


\input{1_introduction}
\input{2_related_work}
\input{3_methodology}
\input{6_results_table}
\input{4_experiments}
\input{5_evaluation}
\input{6_results}
\input{7_conclusion}

\input{AI_disclosure}
\bibliographystyle{IEEEtran}
\bibliography{references}

\end{document}

%% file: 0_abstract.tex
\begin{abstract}
We present TVBC (Time-Varying Biquad Cascade), an interpretable, low-latency speech enhancement model for real-time, on-device assistive hearing. A lightweight neural controller predicts, in real time, the coefficients of a differentiable cascade of 35 second-order IIR filters (biquads), so the model tracks non-stationary noise while keeping a fully interpretable processing chain: every spectral modification is an explicit, adjustable equalizer curve rather than an opaque ``black-box'' transform. Because the biquad cascade carries the signal processing, the controller can be made very small, driving the cascade with only 24k parameters at a 10.7\,ms algorithmic latency, within hearing-aid budgets, and running entirely on-device so that audio never leaves the device. We also expose the suppression-versus-preservation trade-off as an explicit control: it can be set during training through the loss weighting, and adjusted at inference, with no retraining, by mixing the noisy input with the denoised output. On hearing-aid metrics (HASPI/HASQI) the 24k model stays within about 0.02 of DFNet3 (2.3M parameters, almost two orders of magnitude larger) while using about $29\times$ fewer multiply-accumulates, although larger black-box models still lead on reference metrics such as PESQ. We present TVBC as a proof of concept for a compact, interpretable, and controllable denoiser for on-device assistive hearing.
\end{abstract}

%% file: 1_introduction.tex
\section{Introduction}

Deep learning has transformed speech enhancement, but traditional Digital Signal Processing (DSP) remains attractive for low-power, on-device use because it is computationally cheap and interpretable. Classic DSP, however, cannot track dynamic, non-stationary noise without manual tuning. Differentiable DSP (DDSP) \cite{DDSP} narrows this gap by embedding DSP blocks into trainable pipelines, yet most methods remain non-causal or offline. Unconstrained neural models excel at waveform matching but behave as ``black boxes'' and often introduce artifacts that degrade perceived quality. These properties matter most in assistive hearing, where a denoiser needs low latency, on-device operation for privacy, predictable and adjustable behavior, and freedom from synthetic artifacts.

We introduce Time-Varying Biquad Cascade (TVBC), a lightweight, low-latency system for real-time speech enhancement built around these constraints. A neural backbone predicts the time-varying coefficients of a cascade of 35 second-order IIR filters from the input audio, so enhancement is performed by interpretable linear filtering rather than opaque masking. Our contributions are: (i) an interpretable, on-device denoiser built on a real-time, ML-controlled biquad cascade, trained efficiently via a vectorized systolic formulation; (ii) a suppression-versus-preservation control that can be fitted per listener, set during training or adjusted at inference; and (iii) a compact realization in which a small recurrent controller drives the cascade with 24k parameters at 10.7,ms latency, staying close to larger models on hearing-aid metrics.

We evaluate on the Valentini benchmark \cite{valentini} and a harder remixed-noise test, reporting hearing-aid metrics (HASPI/HASQI) alongside PESQ and SIGMOS, and we compare against a retrained DFNet3 \cite{DFNet3} and the tiny black-box models GTCRN \cite{gtcrn} and RNNoise \cite{denoisingRNNNoise}. On hearing-aid metrics the 24k TVBC stays close to models orders of magnitude larger; black-box models lead on reference metrics such as PESQ. TVBC targets a different point that assistive hearing needs: an interpretable, controllable, on-device denoiser.

%% file: 2_related_work.tex
\section{Related Work}

The Differentiable Digital Signal Processing (DDSP) framework was introduced by \cite{DDSP}, using a spectral synthesizer to reconstruct and transfer musical instrument timbre. Differentiable parametric equalizers and biquad filters were proposed independently by several groups at DAFx 2020 \cite{bhattacharyaeq, nercessianeq, differentiableIIR}. In the domain of audio effects, \cite{styletransfer} consolidated these ideas into differentiable implementations of a Parametric Equalizer (PEQ) and a Dynamic Range Compressor, whose biquad coefficient formulas we reuse for our filters. Their work also underpins two widely used libraries in the field\footnote{\url{https://github.com/magenta/ddsp}\label{fn:dasp_repo}}\textsuperscript{,}\footnote{\url{https://github.com/csteinmetz1/dasp-pytorch}}.
Recent work extends DDSP to resource-constrained, real-time edge settings: an ultra-lightweight differentiable DSP vocoder \cite{ddspvocoder} and its refinement for speech enhancement \cite{ddspvocoder2}, which drives the vocoder with a compact network.

One active line of work is the efficient implementation of IIR filters, typically realized as a cascade of biquad filters \cite{differentiablebiquad, differentiableIIR}. Recent work introduced efficient differentiable time-varying all-pole filters \cite{allpoles}, and extended this to direct-form biquads \cite{newbiquad}. These methods optimize the efficiency of backpropagation through the filters rather than real-time inference. To our knowledge there is no prior example of an ML controller driving a real-time, time-varying biquad cascade for denoising. The closest exception is \cite{metaaf}, which uses time-varying FIR filters, but relies on a feedback-loop configuration that limits it to tasks like echo cancellation. \cite{deq} proposes a dynamic equalizer related to our work, but its bi-directional Gate Recurrent Unit (GRU) is non-causal and unsuitable for real-time processing.

For speech denoising, large-scale generative models \cite{SAM, SGMSE} reach very high quality but are non-causal and too expensive on-device. Lightweight real-time models (roughly 1--10M parameters) include DeepFilterNet \cite{DFNet1, DFNet2, DFNet3}, DCCRN \cite{denoisingDCCRN, denoisingSDCCRN, denoisingDCCRN+}, GTCRN \cite{gtcrn}, and others \cite{denoisingDMFNet, denoisingNSNET, denoisingCLCNet, denoisingFullSubNet}; RNNoise \cite{denoisingRNNNoise} is an 85k-parameter hybrid DSP/ML approach. Closest to us, \cite{steinmetzdenoiser} pairs a neural controller with a differentiable DSP denoiser, but targets offline use. Our controller is far smaller than typical neural denoisers (24k parameters). We retrain DFNet3 \cite{DFNet3} as the main reference and compare against the tiny black-box GTCRN \cite{gtcrn} and RNNoise \cite{denoisingRNNNoise}.

Speech enhancement for hearing aids carries its own constraints. Latency must be very low, often under 10\,ms, to avoid disturbing the wearer; power must be low for always-on use; and noise removal has to be balanced against audibility, since over-suppression and processing artifacts are especially harmful for impaired listeners. These needs motivated the Clarity challenges \cite{clarity} and hearing-aid-oriented models such as CLCNet \cite{denoisingCLCNet} and compact low-latency multichannel networks like GCFSnet \cite{gcfsnet}, and they are reflected in evaluation with HASPI \cite{haspi} and HASQI \cite{hasqi} rather than reference-only metrics. TVBC is built for this setting: interpretable, low-latency, adjustable, and small enough to run on-device.

%% file: 3_methodology.tex
\section{Methodology}
\input{3_model_architecture}

Our proposed system is a machine learning pipeline that controls a chain of 35 cascaded biquad filters. The input audio is segmented into non-overlapping 512-sample frames at 48\,kHz, a 10.7\,ms window that fits hearing-aid latency budgets. These frames are processed by two branches, as shown in Figure \ref{Figure:model}. The backbone analyzes the signal in the spectral domain and predicts three control parameters, gain ($g$), quality factor ($q$), and center frequency ($f_0$), for each of the 35 biquads per frame. The cascade filters the frame in the time domain using these time-varying coefficients, and the processed frames are concatenated to reconstruct the output. The backbone's controller is our compact Fast-Weight Programmer (FWP) cell (Section~\ref{sec:fwp}).

\subsection{Machine Learning Backbone}
For each frame, the neural backbone processes the 257-bin magnitude spectrum with two 1D convolutional layers (kernel size 5, stride 2). These layers raise the channel depth to 4 while halving the spectral dimension twice. The flattened result feeds our 2-layer FWP controller of width 32 (Section~\ref{sec:fwp}), followed by a linear projection head with a sigmoid activation that maps to the 105 filter parameters (3 parameters $\times$ 35 filters). The parameters are scaled to physical ranges: gain $[-20, 20]$\,dB and quality factor $[0.1, 2.0]$ for all 35 filters, while the frequency range $[f_\text{min}, f_\text{max}]$ is specific to each filter.

A recurrent controller is well suited here because its temporal memory keeps the predicted coefficients consistent across frames, which prevents frequency-response discontinuities that would otherwise cause audible clicks and pops when filter coefficients change. Because the biquad cascade is parameter-free and carries the signal processing, the controller can be kept very small: the full model totals 24\,k parameters, with the budget dominated by the controller and the projection head.

\subsection{FWP Controller}
\label{sec:fwp}
A controller for a time-varying filter has to emit a smoothly varying parameter trajectory at every frame. A standard recurrent unit such as a GRU could fill this role; we instead use a more compact cell, which we call the FWP controller, that can be viewed as a subset of a GRU. Where a GRU applies three gated input projections to the wide flattened-spectrum input, our cell applies a single input projection and carries its recurrent state as a leaky integrator. Two properties follow and motivate this choice for filter control. First, the single projection makes the cell smaller at matched width, which keeps the controller a small fraction of an already tiny model. Second, the leaky-integrator state bounds how fast the emitted coefficients can change between frames, so the predicted equalizer curves vary smoothly by construction and avoid the abrupt jumps that cause audible clicks. The cell belongs to the fast-weight programmer family \cite{fastweights}, where a slow network writes the parameters of a fast one; ours is a fully classical reduction of a quantum fast-weight programmer \cite{qfwp_origin}, with the recurrent state playing the role of the fast weights and no quantum hardware involved.

At each time step $t$, given the backbone features $\mathbf{x}_t$, a slow programmer produces $\mathbf{h}_t = \mathrm{ReLU}(\mathbf{W}_p \mathbf{x}_t)$. A fast-weight state $\boldsymbol{\Phi}_t$ is then updated as a leaky integrator, and a re-uploaded copy of the input is read out through a small programmed network $F$:
\begin{align}
\boldsymbol{\Phi}_t &= \boldsymbol{\gamma} \odot \boldsymbol{\Phi}_{t-1} + \mathbf{W}_{\phi}\, \mathbf{h}_t, \\
\mathbf{d}_t &= \tanh(\mathbf{W}_d\, \mathbf{h}_t), \\
\mathbf{y}_t &= F\!\left([\,\mathbf{d}_t\,;\,\boldsymbol{\Phi}_t\,]\right),
\end{align}
where $\boldsymbol{\gamma} \in (0,1)$ is a per-dimension decay (default $0.9$, optionally learnable), $\odot$ is the elementwise product, and $F$ is a two-layer MLP whose input is the concatenation of $\mathbf{d}_t$ and $\boldsymbol{\Phi}_t$. The per-dimension decay $\boldsymbol{\gamma}$ regulates the smoothness that the leaky state imposes on the emitted coefficients.

\subsection{Time Varying IIR Filter Cascade}
The filtering stage employs a differentiable cascade of second-order IIR filters (biquads). For frame $n$, the $k$-th biquad is defined by the transfer function:
\begin{equation}
    H^{(k, n)}(z) = \frac{b_0^{(k, n)} + b_1^{(k, n)} z^{-1} + b_2^{(k, n)} z^{-2}}{1 + a_1^{(k, n)} z^{-1} + a_2^{(k, n)} z^{-2}}, \label{EQ:biquad}
\end{equation}
where $b^{(k, n)}_i$ and $a^{(k, n)}_i$ are the feedforward and feedback coefficients predicted for the whole frame. We parameterize the filters using three properties: gain ($g$), quality factor ($q$), and center or cutoff frequency ($f_0$). We map these attributes to the filter coefficients $\textbf{a}, \textbf{b}$ using the same formulae as in \cite{styletransfer} to obtain the desired frequency response shapes. Because these RBJ formulae apply $a_0$-normalization with a bounded quality factor, both poles of every section lie strictly inside the unit circle for any predicted $(f_0, q, g)$, so stability is structural. Empirically the worst-case pole radius across the $6.8$M biquad sections emitted on the 824-file test set is $0.9998$, so the cascade cannot become unstable.

Our chain consists of a low-frequency suppression filter, 33 band-pass resonant filters, and a high-frequency roll-off. To constrain the optimization, each band $k$ has its center frequency $f_0^{(k)}$ bounded to a per-band interval $[f_\text{min}^{(k)}, f_\text{max}^{(k)}]$. The two bounding filters are restricted to $[20, 60]$\,Hz and $[12000, 22000]$\,Hz, which biases them toward suppressing low-frequency rumble and high-frequency hiss. The 33 inner bands follow a hybrid spacing: their intervals tile the spectrum linearly with a width of about 50\,Hz up to 1\,kHz, to resolve the fundamental frequencies of speech, and then widen geometrically above 1\,kHz to cover the higher formants with fewer bands. The choice of 35 sections balances spectral resolution against the per-sample cost of the cascade; the band count and spacing are fixed design choices, and a systematic ablation over them is left to future work.

\subsection{IIR Filtering Implementation}

We implement the time-varying filtering of Equation \ref{EQ:biquad} frame-by-frame using Direct Form I \cite[pp.~393--400]{dspbook}:
\begin{align}
w^{(k,n)}[t] &= \sum_{i=0}^{2} b_i^{(k, n)} x^{(k)}[t-i]\label{EQ:fir}, \\
y^{(k,n)}[t] &= w^{(k,n)}[t] - \sum_{i=1}^{2} a_i^{(k, n)} y^{(k)}[t-i]\label{Eq:allpoles},
\end{align}
where $x^{(k)}[t]$, $y^{(k)}[t]$, and $w^{(k)}[t]$ are the $t$-th time sample of the $n$-th frame of the input, output, and intermediate state for the $k$-th filter, respectively. Note that the input to the first filter is the original audio ($x^{(0)}[t] = x[t]$), and subsequent filters take the previous output ($x^{(k)}[t] = y^{(k-1)}[t]$). We compute Equation \ref{Eq:allpoles} using the recursive implementation of an all-poles filter from \cite{allpoles}\footnote{\url{https://github.com/DiffAPF/torchlpc}}. We pass the last 2 samples of both $x^{(k)}$ and $y^{(k)}$ to the next frame to handle the state correctly.

A naive sequential computation of the $K$-biquad cascade over $N$ frames requires a nested loop of depth $K \times N$. To optimize this, we adapt the systolic processing approach \cite{systolic} into a vectorized tensor formulation, reducing the depth to $N+K-1$. We construct time-shifted coefficient matrices $\tilde{\mathbf{A}}, \tilde{\mathbf{B}} \in \mathbb{R}^{K \times (N+K-1) \times 3}$. For a generic matrix $\tilde{\mathbf{C}} \in \{\tilde{\mathbf{A}}, \tilde{\mathbf{B}}\}$, the parameter sequence $\mathbf{c}_n^{(k)} \in \mathbb{R}^3$ of the $k$-th filter is shifted forward by $k$ zero-padded steps. Concurrently, we define a parallel input vector $\mathbf{X}_n \in \mathbb{R}^K$:
\begin{equation}
\begin{split}
&\tilde{\mathbf{C}} =
\begin{bmatrix}
\mathbf{c}_0^{(0)} & \mathbf{c}_1^{(0)} & \dots & \mathbf{c}_{N-1}^{(0)} & \mathbf{0} & \dots \\
\ddots & \ddots & \ddots & \ddots & \ddots & \ddots \\
\dots & \mathbf{0} & \mathbf{c}_0^{(K-1)} & \mathbf{c}_1^{(K-1)} & \dots & \mathbf{c}_{N-1}^{(K-1)}
\end{bmatrix}, \\
&\mathbf{X}_n = [\mathbf{x}_n, \mathbf{y}_{n-1}^{(0)}, \dots, \mathbf{y}_{n-1}^{(K-2)}]^\top.
\end{split}
\end{equation}
The diagonal of $\tilde{\mathbf{C}}$ holds, at step $n$, the coefficients that the $k$-th filter needs for the frame it is currently processing: shifting filter $k$ forward by $k$ steps lines up each filter one stage behind the previous one, exactly as data flows through a systolic array. At any execution step $n \in [0, N+K-2]$, the $n$-th column slices of $\tilde{\mathbf{A}}$ and $\tilde{\mathbf{B}}$ therefore align the coefficients required to process all $K$ filters at once, and feeding these slices and $\mathbf{X}_n$ into Equations \ref{EQ:fir} and \ref{Eq:allpoles} evaluates the whole cascade for that step with parallel matrix operations rather than a $K \times N$ deep recurrence. The speed-up depends only on the cascade depth $K$ and the number of frames $N$, not on the frame length $L$ (each step still filters all $L$ samples of a frame), so it is independent of the analysis window. The price is a $K{-}1$ frame shift during training, which is purely an artifact of the vectorized layout; at inference we run the standard serial implementation, which has no such shift and keeps the algorithmic latency at one frame (10.7\,ms).

\subsection{Weights Initialization}
To improve training stability, we initialize the final linear layer's gain parameters near 0 dB (with minor noise) so the model begins in an ``all-pass'' state. Standard random initialization often caused overly aggressive initial frequency responses, trapping the model in poor local minima that suppressed the entire signal or wasting dozens of epochs just learning to pass the signal through. Our approach prevents these pitfalls and significantly accelerates convergence.

%% file: 3_model_architecture.tex
\begin{figure*}[!ht]
    \centering
    \includegraphics[width=\textwidth]{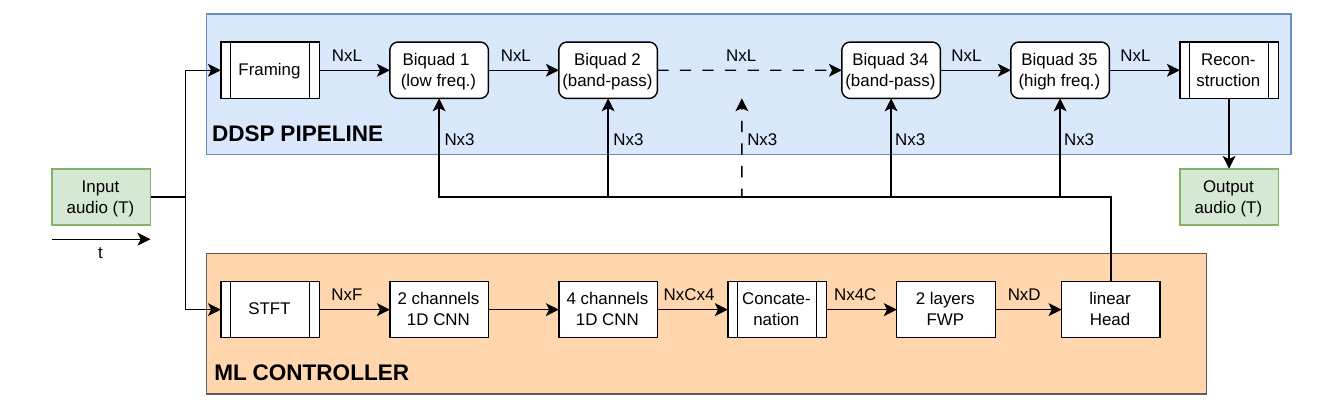}
    \caption{Model architecture. $T$ samples, $N$ frames, $L{=}512$ frame length (10.7\,ms at 48\,kHz), $F{=}257$ frequency bins, $C$ features per channel after the two convolutions, $D{=}32$ FWP controller width.}
    \label{Figure:model}
\end{figure*}

%% file: 6_results_table.tex
\begin{table}[t!]
    \renewcommand{\arraystretch}{1.2}
    \setlength{\tabcolsep}{6pt}
    \caption{Standard 824-file Valentini test (HASPI/HASQI at the moderate audiogram). TVBC against RNNoise, the GTCRN-DNS3 checkpoint, and the much larger DFNet3.}
    \label{tab:standard}
    \centering
    \begin{tabular}{lccc@{\hskip 10pt}c}
    \toprule
    \textbf{Metric} & \textbf{RNNoise} & \textbf{GTCRN-DNS3} & \textbf{TVBC} & \textbf{DFNet3} \\
    \midrule
    Params $\downarrow$ & 85\,k & \textbf{24\,k} & \textbf{24\,k} & 2.31\,M \\
    \midrule
    HASPI $\uparrow$    & 0.824 & \textbf{0.830} & 0.815 & 0.818 \\
    HASQI $\uparrow$    & 0.439 & 0.459 & 0.441 & \textbf{0.462} \\
    PESQ $\uparrow$     & 2.11 & 2.52 & 2.06 & \textbf{3.05} \\
    SIGMOS $\uparrow$   & 2.85 & 3.11 & 2.59 & \textbf{3.37} \\
    eSTOI $\uparrow$    & 0.78 & 0.81 & 0.79 & \textbf{0.86} \\
    SI-SDR $\uparrow$   & 12.3 & 15.5 & 12.8 & \textbf{18.5} \\
    \bottomrule
    \end{tabular}
\end{table}

\begin{table}[t!]
    \centering
    \small
    \setlength{\tabcolsep}{4pt}
    \caption{Harder-noise test: 824 files remixed at fixed SNR (moderate audiogram).}
    \label{tab:hard}
    \resizebox{\columnwidth}{!}{%
    \begin{tabular}{clcccc}
    \toprule
    SNR (dB) & Metric & RNNoise & GTCRN-DNS3 & TVBC & DFNet3 \\
    \midrule
    $-5$ & PESQ $\uparrow$  & 1.50 & 1.73 & 1.27 & \textbf{1.93} \\
    $-5$ & HASPI $\uparrow$ & 0.752 & \textbf{0.783} & 0.690 & 0.728 \\
    $-5$ & HASQI $\uparrow$ & 0.300 & \textbf{0.334} & 0.275 & 0.302 \\
    \midrule
    $0$  & PESQ $\uparrow$  & 1.71 & 2.06 & 1.51 & \textbf{2.39} \\
    $0$  & HASPI $\uparrow$ & 0.806 & \textbf{0.819} & 0.791 & 0.814 \\
    $0$  & HASQI $\uparrow$ & 0.352 & \textbf{0.381} & 0.342 & 0.367 \\
    \midrule
    $5$  & PESQ $\uparrow$  & 1.97 & 2.42 & 1.82 & \textbf{2.81} \\
    $5$  & HASPI $\uparrow$ & 0.824 & 0.824 & 0.810 & \textbf{0.825} \\
    $5$  & HASQI $\uparrow$ & 0.381 & \textbf{0.401} & 0.371 & 0.393 \\
    \bottomrule
    \end{tabular}%
    }
\end{table}

\begin{table}[t]
    \centering\small
    \setlength{\tabcolsep}{4pt}
    \caption{HASPI / HASQI by audiogram severity (standard test): mild, moderate, and moderately severe hearing loss.}
    \label{tab:severity}
    \resizebox{\columnwidth}{!}{%
    \begin{tabular}{lccc@{\hskip 10pt}ccc}
    \toprule
     & \multicolumn{3}{c}{HASPI $\uparrow$} & \multicolumn{3}{c}{HASQI $\uparrow$} \\
    \cmidrule(lr){2-4}\cmidrule(lr){5-7}
    Model & mild & mod & m-sev & mild & mod & m-sev \\
    \midrule
    RNNoise          & 0.875 & 0.824 & 0.680 & 0.532 & 0.439 & 0.311 \\
    GTCRN-DNS3       & \textbf{0.880} & \textbf{0.830} & \textbf{0.708} & 0.544 & 0.459 & \textbf{0.333} \\
    TVBC (24\,k)      & 0.872 & 0.815 & 0.681 & 0.536 & 0.441 & 0.310 \\
    DFNet3 (2.31\,M) & 0.872 & 0.818 & 0.700 & \textbf{0.554} & \textbf{0.462} & 0.332 \\
    \bottomrule
    \end{tabular}%
    }
\end{table}

\begin{table}[t]
    \centering\small
    \setlength{\tabcolsep}{4pt}
    \caption{Effect of training data on TVBC (Valentini test set): Valentini-56 only vs Valentini-56 + DNS.}
    \label{tab:dns}
    \resizebox{\columnwidth}{!}{%
    \begin{tabular}{llcccc}
    \toprule
    test & data & PESQ $\uparrow$ & HASPI $\uparrow$ & HASQI $\uparrow$ & MOS-Noise $\uparrow$ \\
    \midrule
    standard  & Valentini & \textbf{2.06} & 0.804 & 0.431 & \textbf{3.16} \\
    standard  & +DNS      & 2.05 & \textbf{0.807} & \textbf{0.446} & 2.98 \\
    \midrule
    hard $-5$ & Valentini & 1.32 & 0.640 & 0.252 & \textbf{2.62} \\
    hard $-5$ & +DNS      & \textbf{1.33} & \textbf{0.694} & \textbf{0.288} & 2.35 \\
    \midrule
    hard $0$  & Valentini & 1.55 & 0.767 & 0.322 & 2.79 \\
    hard $0$  & +DNS      & 1.54 & \textbf{0.784} & \textbf{0.345} & 2.60 \\
    \midrule
    hard $5$  & Valentini & \textbf{1.83} & 0.804 & 0.366 & \textbf{3.03} \\
    hard $5$  & +DNS      & 1.81 & \textbf{0.809} & \textbf{0.377} & 2.81 \\
    \bottomrule
    \end{tabular}%
    }
\end{table}

\begin{table}[t]
    \centering\small
    \caption{Computational cost: multiply-accumulates per second of audio. TVBC and DFNet3 measured with shape-based hooks plus analytical FFT and biquad terms; GTCRN~\cite{gtcrn} (16\,kHz) and RNNoise~\cite{denoisingRNNNoise} as reported by their authors.}
    \label{tab:cost}
    \begin{tabular}{lccc}
    \toprule
    Model & Params $\downarrow$ & MAC/s $\downarrow$ & GFLOP/s $\downarrow$ \\
    \midrule
    TVBC                 & \textbf{24\,k}   & \textbf{11.3\,M}       & \textbf{0.023} \\
    RNNoise             & 85\,k   & $\sim$20\,M   & $\sim$0.040 \\
    GTCRN-DNS3          & \textbf{24\,k}   & 39.6\,M   & 0.079 \\
    DFNet3              & 2.31\,M & 325.7\,M      & 0.651 \\
    \bottomrule
    \end{tabular}
\end{table}

%% file: 4_experiments.tex
\section{Experiments}
We evaluate TVBC on a speech denoising task, comparing it against the retrained DeepFilterNet3 (DFNet3) \cite{DFNet3} and the tiny black-box models GTCRN \cite{gtcrn} and RNNoise \cite{denoisingRNNNoise}.

\subsection{Dataset}
We train on the 56-speaker Valentini-Botinhao noisy speech dataset \cite{valentini}. It is small by modern standards but remains a standard benchmark for comparing architectures under a fixed data budget. The dataset contains 19 hours of clean speech mixed with various noises. We reserve 12 speakers for validation and train on the remaining 44.

We obtain pure noise tracks as the difference between the clean and noisy pairs and use them for augmentation. Following the DeepFilterNet recipe, we dynamically remix random speech and noise files at each iteration with a signal-to-noise ratio (SNR) drawn from $\{-5, 0, 5, 10, 20, 40, 100\}$\,dB; the 100\,dB case teaches the models to leave clean audio untouched. For the deployed compact models we widen the noise pool with samples from the DNS corpus \cite{dns} to improve generalization to non-stationary noise. This expansion is meant to increase noise diversity.

We use two test sets. The first is the standard Valentini test set of 824 files from two held-out speakers, paired with clean references. The second is a harder test built by remixing those files at SNRs of $-5$, $0$, and $5$\,dB, which stresses the lower-SNR, more non-stationary regime relevant to hearing aids.

\subsection{Training Setup}
We retrain DFNet3 \cite{DFNet3} from scratch under the same data budget, following the official hyperparameters\footnote{\url{https://github.com/Rikorose/DeepFilterNet}} with a batch size of 64. Trained to convergence it represents a strong baseline that leads TVBC on reference metrics. GTCRN and RNNoise are used as released pretrained checkpoints, trained at 16\,kHz on different corpora; they are therefore external references rather than data-matched baselines (GTCRN is resampled to and from 16\,kHz for evaluation), and comparisons to them should be read with that caveat.

For TVBC we minimize a multi-scale logarithmic spectral distance plus a time-domain mean-squared-error term, with the time-domain term scaled by $5 \times 10^4$ to balance the two. This weight also sets the suppression-versus-preservation balance, which we revisit in Section~\ref{sec:control}. We use a batch size of 64 and Adam \cite{adam, pytorch} with a fixed learning rate of $10^{-3}$ ($\beta_1 = 0.9$, $\beta_2 = 0.999$, $\epsilon = 10^{-8}$), training for up to 100 epochs and keeping the best validation checkpoint.

%% file: 5_evaluation.tex
\subsection{Evaluation Metrics}
Because our target application is assistive hearing, we lead with hearing-aid metrics. HASPI \cite{haspi} predicts speech intelligibility and HASQI \cite{hasqi} predicts speech quality for a listener with a given audiogram. Both compares the processed signal to the clean reference through an auditory-periphery model, and are scaled 0--1. We compute them with standard audiograms at three severities (mild, moderate, and moderately severe) and report the moderate setting unless stated otherwise.

We complement these with standard speech-enhancement metrics. PESQ \cite{metricpesq} (1--5) measures perceptual quality, and SIGMOS \cite{metricsigmos}, a reference-free DNSMOS-style estimator \cite{metricdnsmos}, reports MOS-Signal, MOS-Noise, and MOS-Overall (1--5). For completeness we also report eSTOI \cite{metricestoi} (intelligibility, 0--1) and SI-SDR \cite{metricsisdr} (time-domain distortion, dB). Higher is better for every metric we report.

%% file: 6_results.tex
\section{Results}
Tables~\ref{tab:standard} and~\ref{tab:hard} report the standard and harder-noise tests. DFNet3 is a much larger model from a different class (2.31\,M parameters); TVBC targets a compact, interpretable, and controllable design point, and on the hearing-aid metrics it stays close to that model at a small fraction of the size.

\subsection{Reference Quality}
On reference metrics the black-box models lead. DFNet3, a mask-based model almost two orders of magnitude larger, reaches PESQ\,$3.05$ on the standard test and is ahead at every SNR on the harder test, and the compact GTCRN-DNS3 also beats our model on PESQ ($2.52$ vs $2.06$), though at about $3.5\times$ the MAC/s of TVBC (Table~\ref{tab:cost}). These models reconstruct magnitude and phase with an opaque mask. TVBC is restricted to interpretable linear filtering and does not reconstruct phase, so it trades these waveform-matching metrics for interpretability, control, and a smaller on-device footprint.

\subsection{Hearing-Aid Metrics}
The picture changes on HASPI and HASQI, the relevant metrics for hearing aids. On the standard test (Table~\ref{tab:standard}) TVBC (24\,k) lands within about $0.02$ of DFNet3 (2.31\,M, two orders of magnitude larger) on both indices, and in the same band as GTCRN-DNS3. Table~\ref{tab:severity} breaks this down by audiogram: in mild, moderate, and moderately severe hearing loss the four models stay within about $0.02$ HASPI of one another, with GTCRN-DNS3 slightly ahead and DFNet3 best on HASQI at the milder losses, while TVBC stays within $0.01$--$0.02$ of the best at every severity despite being the smallest model. The harder-noise test (Table~\ref{tab:hard}) is more informative: there GTCRN-DNS3 and DFNet3 lead HASPI at low SNR, with TVBC a few points behind at a fraction of their size.

\subsection{Parameter Efficiency}
Because the parameter-free biquad cascade carries the signal processing, the neural controller can be kept tiny: in TVBC it is only a small fraction of a model that totals 24\,k parameters and $11.3$\,M\,MAC/s (Table~\ref{tab:cost}), roughly two orders of magnitude smaller than the mask-based DFNet3. Putting the modelling burden on an interpretable filter rather than a large network is what makes this footprint possible.

\subsection{Interpretability}
A defining property of TVBC is that its action is fully readable. Figure~\ref{Figure:spectrograms} plots the predicted frequency response over time for a clip with non-stationary noise. During the initial noise-only segment the model applies a strong broadband attenuation; when speech begins, it opens up the speech bands and keeps suppressing the rest. The transition is smooth because the controller state changes gradually, which avoids the coefficient jumps that would otherwise create clicks. Every decision the model makes is an explicit equalizer curve, which a black-box mask cannot offer.
\input{6_picture}

\subsection{Controllability}
\label{sec:control}
How much noise to remove is a personal choice: for a hearing-aid wearer, too little suppression leaves disturbing noise, while too much muffles speech and harms audibility. A denoiser for this setting should therefore let the suppression-versus-preservation balance be adjusted, ideally per listener and even per environment. A defining property of TVBC is that it exposes this balance as an explicit control, in two ways, which we quantify on the hard $-5$\,dB test.

\begin{figure}[t]
    \centering
    \includegraphics[width=\columnwidth]{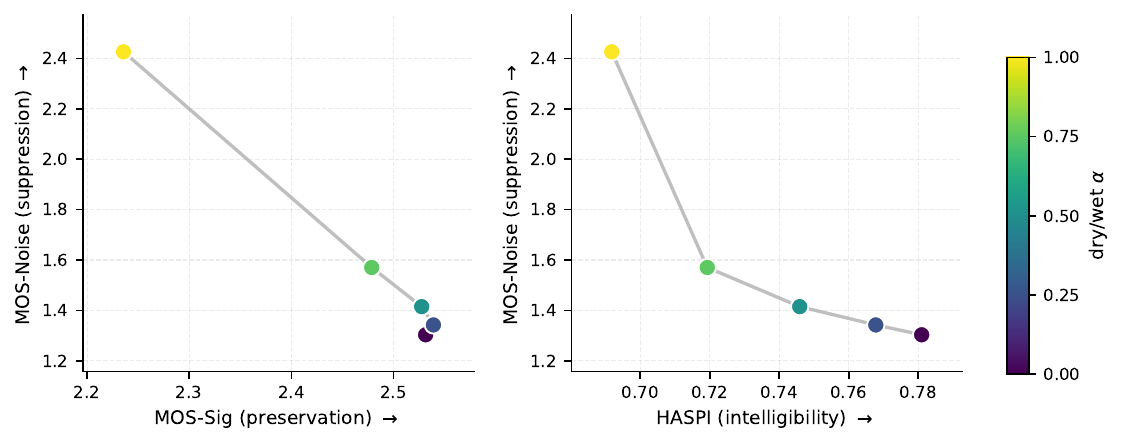}
    \caption{Inference-time control (TVBC, hard $-5$\,dB). Blending the noisy input and the denoised output with $\alpha\in[0,1]$ (colour) trades speech preservation (MOS-Sig, left) and intelligibility (HASPI, right) for noise suppression (MOS-Noise), with no retraining.}
    \label{fig:control_rt}
\end{figure}

A single deployed model can be made more or less aggressive by blending its input and output. We combine the noisy input $x$ and the denoised output $\hat{x}$ with one coefficient $\alpha \in [0,1]$, $y = (1-\alpha)\,x + \alpha\,\hat{x}$: at $\alpha=0$ the wearer hears the unprocessed signal, at $\alpha=1$ full denoising, and intermediate values apply partial suppression. Increasing $\alpha$ from 0 to 1 raises noise suppression (MOS-Noise $1.30\rightarrow2.43$) while gradually giving up speech preservation (MOS-Signal $2.53\rightarrow2.24$) and intelligibility (HASPI $0.78\rightarrow0.69$), tracing the trade-off curve in Figure~\ref{fig:control_rt}. Since this is one scalar applied at run time, a wearer or audiologist can set it on the fly without touching the model.

The same balance can be chosen before deployment through the weight of the time-domain loss term, which controls how hard the model is pushed toward suppression. Raising it from $1{,}000$ to $200{,}000$ trains a model that suppresses more and is more intelligible (MOS-Noise $2.30\rightarrow2.44$, HASPI $0.62\rightarrow0.71$) at the cost of speech preservation (MOS-Signal $2.42\rightarrow2.23$), so variants can be tuned to different listener profiles.

\subsection{Computational Cost}
TVBC is cheap to run. Table~\ref{tab:cost} reports the cost in multiply-accumulates per second of audio. TVBC needs about $11.3$\,M\,MAC/s ($0.023$\,GFLOP/s), roughly $29\times$ fewer than DFNet3 ($325.7$\,M\,MAC/s) and the lowest of the compared models, below the tiny black-box baselines RNNoise ($\sim$20\,M) and GTCRN ($39.6$\,M). The cost is dominated by the interpretable biquad cascade (about three quarters), with the neural controller, convolutions, and a single analysis STFT making up the rest; there is no inverse transform, since the model filters in the time domain. With the $10.7$\,ms algorithmic latency, this keeps TVBC within an on-device, real-time budget.

\subsection{Effect of Training Data}
\label{sec:dns}
We test whether a larger, more diverse training corpus helps, holding the model fixed (TVBC) and changing only the data: Valentini-56 alone versus Valentini-56 plus the DNS corpus (Table~\ref{tab:dns}). On the easy standard test the two are within noise. Under hard noise the diverse corpus helps the hearing-aid metrics, most at the lowest SNR ($-5$\,dB: HASPI $0.640\rightarrow0.694$, HASQI $0.252\rightarrow0.288$), while SIGMOS and SI-SDR change little or drop slightly because the expanded model suppresses less aggressively. The gain is concentrated exactly in the hard, non-stationary regime that matters for assistive hearing.

%% file: 6_picture.tex
\begin{figure}[!ht]
    \centering
    \includegraphics[width=0.45\textwidth]{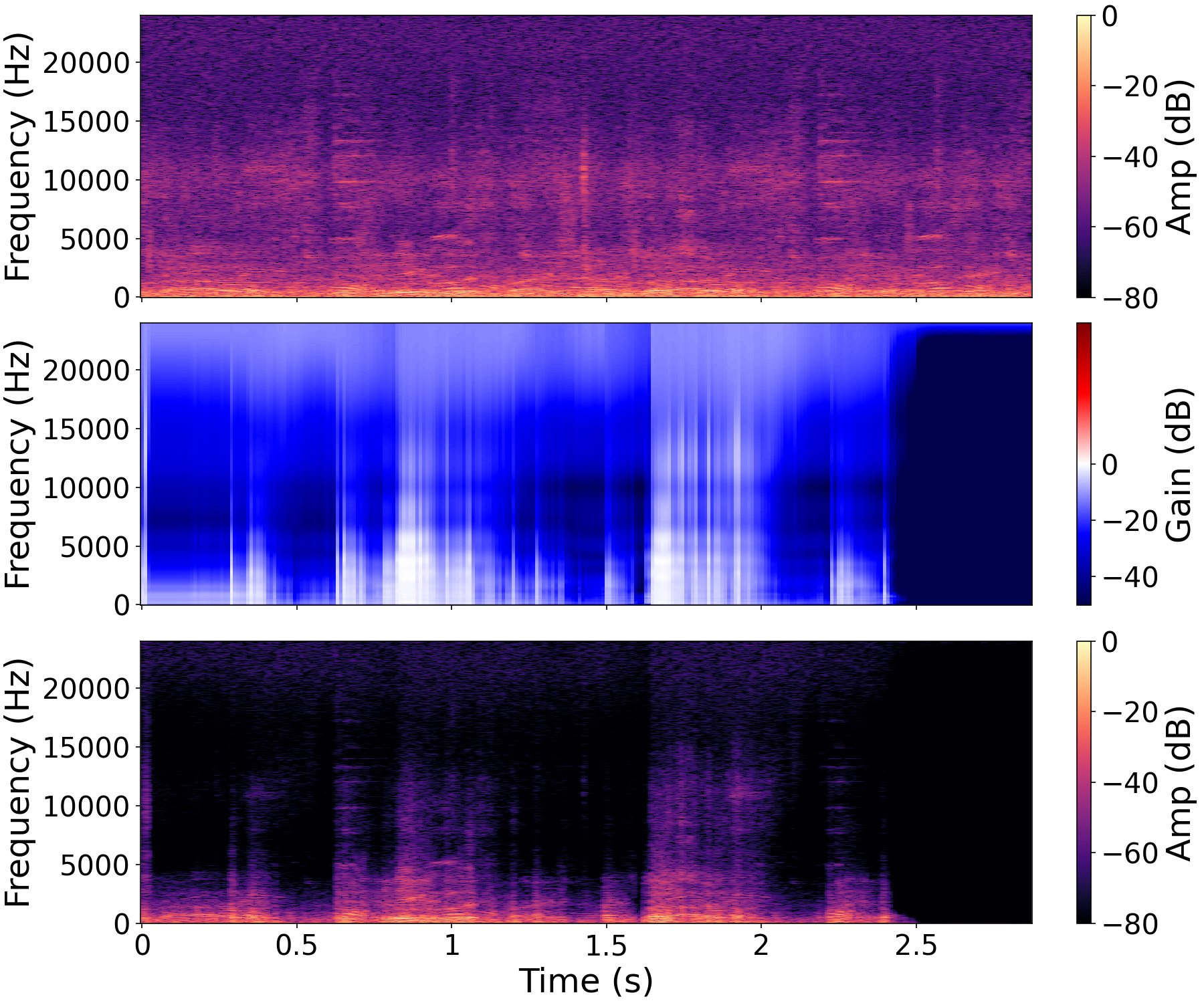}
    \caption{TVBC on a test utterance under heavy non-stationary noise (hard $-5$\,dB). Top to bottom: noisy input, predicted time-varying EQ gain, denoised output.}
    \label{Figure:spectrograms}
\end{figure}

%% file: 7_conclusion.tex
\section{Conclusion and Future Work}
TVBC and black-box denoisers such as DFNet3 rest on different paradigms. A deep STFT-domain masker predicts complex weights to reconstruct magnitude and phase, which makes it strong on waveform-matching metrics but opaque and prone to synthesis artifacts. TVBC instead maps neural outputs to physically constrained, time-varying biquad parameters and filters the signal in the time domain. This inductive bias is a genuine limitation: it cannot reconstruct phase or perform non-linear separation, and larger black-box models lead the reference metrics. In return it provides a fully interpretable control surface, stable linear processing without neural synthesis artifacts, and a small, low-latency footprint that runs entirely on-device, keeping audio local and preserving user privacy. For assistive hearing this trade is attractive: a 24\,k TVBC stays within about $0.02$ of a 2.31\,M model across audiogram severities while using roughly $29\times$ fewer multiply-accumulates, the FWP controller compresses the model by more than an order of magnitude at little quality cost, and the suppression-versus-preservation balance is exposed as a control at both training and inference time. As a proof of concept, it shows that interpretable, controllable, on-device denoising is viable for assistive hearing, even if larger black-box models remain ahead on raw reference metrics.

The main limitations are the purely linear filtering, which does not reconstruct phase and cannot perform the non-linear separation that complex masks can, and an evaluation that is monaural and on simulated mixtures: we rely on HASPI, HASQI, and SIGMOS as proxies rather than subjective listening tests, and we do not test real hearing-aid acoustics or hearing-impaired listeners. Future work will add listening tests with hearing-impaired participants under realistic acoustics, ablate the number and spacing of filter bands, explore lightweight per-listener adaptation on top of the interpretable control surface, train on more diverse corpora to narrow the reference-metric gap, measure on-device real-time factor and power, and extend the controller to multi-channel processing.

%% file: AI_disclosure.tex
\section*{AI Disclosure}
The authors used Claude (Anthropic) to help conduct experiments and to polish the writing of this manuscript. The model architecture and its definition are the authors' own work.